# Composition of Near-Earth Asteroid 2008 EV5: Potential target for Robotic and Human Exploration

Vishnu Reddy[1]
Department of Space Studies, University of North Dakota, Grand Forks, USA.
Max Planck Institute for Solar System Research, Katlenburg-Lindau, Germany.
Email: reddy@space.edu

Lucille Le Corre[1]
Max Planck Institute for Solar System Research, Katlenburg-Lindau, Germany.

Michael Hicks
Jet Propulsion Laboratory, California Institute of Technology, 4800 Oak Grove Drive, Pasadena, CA 91109, USA.

Kenneth Lawrence
Jet Propulsion Laboratory, California Institute of Technology, 4800 Oak Grove Drive, Pasadena, CA 91109, USA.

Bonnie J. Buratti
Jet Propulsion Laboratory, California Institute of Technology, 4800 Oak Grove Drive, Pasadena, CA 91109, USA.

Paul A. Abell[1]
Astromaterials Research & Exploration Science Directorate, NASA Johnson Space Center, Mail Code KR, 2101 NASA Parkway, Houston, TX 77058-3696, USA.

Michael J. Gaffey[1]
Department of Space Studies, University of North Dakota, Grand Forks, ND 58202, USA.

Paul S. Hardersen[1]
Department of Space Studies, University of North Dakota, Grand Forks, ND 58202, USA.

[1]Visiting Astronomer at the Infrared Telescope Facility, which is operated by the University of Hawaii under Cooperative Agreement no. NNX-08AE38A with the National Aeronautics and Space Administration, Science Mission Directorate, Planetary Astronomy Program.

Pages: 19
Figures: 6
Tables: 1
**Proposed Running Head:** Composition of Asteroid 2008 EV5

**Editorial correspondence to:**
Vishnu Reddy
Max-Planck Institute for Solar System Research,
37191 Katlenburg-Lindau,
Germany.
+49-5556-979-550, reddy@mps.mpg.de

48  **Abstract**

49  We observed potentially hazardous asteroid (PHA) 2008 EV5 in the visible (0.30-0.92

50  μm) and near-IR (0.75-2.5 μm) wavelengths to determine its surface composition. This

51  asteroid is especially interesting because it is a potential target for two sample return

52  mission proposals (Marco Polo-R and Hayabusa-2) and human exploration due to its low

53  delta-v for rendezvous. The spectrum of 2008 EV5 is essentially featureless with

54  exception of a weak 0.48-μm spin-forbidden $Fe^{3+}$ absorption band. The spectrum also has

55  an overall blue slope. The albedo of 2008 EV5 remains uncertain with a lower limit at

56  0.05 and a higher end at 0.20 based on thermal modeling. The Busch et al. (2011) albedo

57  estimate of 0.12±0.04 is consistent with our thermal modeling results. The albedo and

58  composition of 2008 EV5 are also consistent with a C-type taxonomic classification

59  (Somers et al. 2008). The best spectral match is with CI carbonaceous chondrites similar

60  to Orgueil, which also have a weak 0.48-μm feature and an overall blue slope. This 0.48-

61  μm feature is also seen in the spectrum of magnetite. The albedo of CI chondrites is at the

62  lower limit of our estimated range for the albedo of 2008 EV5.

63
64
65
66
67
68
69
70

**Introduction**

Potentially hazardous asteroid 2008 EV5 was discovered by the Catalina Sky Survey (Larson et al. 2006) on March 4$^{th}$, 2008. The asteroid was the target of detailed radar investigation (Busch et al. 2011) when it made a close flyby of the Earth in December 2008. Radar observations indicated a retrograde motion (rotation period 3.725±0.001 h) with a diameter of 400±50 m and an oblate spheroidal shape (Busch et al 2011). 2008 EV5 also has a prominent equatorial ridge similar to (66391) 1999 KW4 (Ostro et al. 2006). Radar albedo (0.29±0.009) and optical albedo (0.12±0.04) measurements are consistent with a rocky or a stony/iron composition (Busch et al 2011). Ground-based visible and IR observations of 2008 EV5 show a spectral shape similar to a C- or X-type (Somers et al., 2008; Reddy, 2009).

2008 EV5 is also a potential target for robotic and human exploration mission due its low delta-v for rendezvous (Benner et al 2010; Landau and Strange 2011). The asteroid is a back-up target for the European Space Agency's proposed Marco Polo-R mission and the Japanese Space Agency's Hayabusa-2 mission, which both aim to bring back a sample of an asteroid. For human exploration, Landau and Strange (2011) explored the possibility of a 360-day mission to 2008 EV5 in Dec. 2022 and a 540-day mission in Dec. 2023 at a delta-v of 4.728 km/s and 1.977 km/s, respectively. Given 2008 EV5's potential as a future robotic and human exploration mission target, we conducted detailed mineralogical analysis of the asteroid in order to constrain its surface composition and taxonomic type.

**Observation and Data Reduction**

Visible wavelength (0.30-0.92 μm; R~500) spectral observations of 2008 EV5 were obtained at the Palomar Mountain 200-inch telescope equipped with a facility dual-channel long-slit CCD spectrometer (the "Double-Spec" or DBSP; Oke and Gunn 1982). Using DBSP, the night sky and object were first imaged on the slit before being split by a dichroic filter into blue and red beams. These beams were then dispersed and reimaged with individual grating and camera set-ups. In addition to the target asteroid, each night we obtained spectra of solar analog stars over a wide range of airmass. Wavelength calibrations used arc-lamp exposures and flat-fields were taken using the illuminated dome. Throughout our campaign we used a 6 arcsec wide slit and kept the tailpiece of the telescope rotated to match the parallactic angle. The spectral data reduction was accomplished using IRAF and custom-built code described in (Hicks & Buratti, 2004). The individual asteroid spectra were ratioed to the average of each set. Any changes in slope or flux in the ratios would suggest problems with differential refraction or changing extinction. Following this quality test, no spectral frames were removed from the sample. Similarly, we were able to cross-reference our solar comparison stars taken on various nights with the well-accepted solar analog 16 Cyg B.

Near-infrared (NIR) observations (0.75-2.5 μm; R~150) of 2008 EV5 were obtained using the SpeX instrument on the NASA IRTF (Rayner et al. 2003) on December 20$^{th}$, 2008. Apart from the asteroid, local standard star SAO155420 and solar analog star SAO93936 were also observed for telluric and solar continuum corrections, respectively. The weather conditions at the time of observation were not ideal and the high airmass (1.7-1.9) of the asteroid contributed to the low SNR spectrum. SpeX prism data were processed using the IDL-based Spextool provided by the NASA IRTF (Cushing et al. 2004). The asteroid data were corrected for telluric absorption bands using local standard star observations and ratioed to a solar analog star for solar continuum calibration. Detailed description of data reduction procedure is available in Reddy et al. (2012). Observational circumstances for both visible and IR data are presented in Table 1.

**Analysis**

*Visible Spectrum*

The asteroid was observed on two nights (Dec. 27$^{th}$, 2008; Jan. 18$^{th}$, 2009). Fig. 1 shows the visible spectrum of 2008 EV5 obtained on Dec. 12$^{th}$, 2008. The data are normalized to unity at 0.55 μm. During the first night the asteroid was brighter (13.2 V. Mag) and higher SNR spectrum could be obtained as is evident in the low point-to-point scatter. The second night the asteroid was fainter (16.2 V. Mag), resulting in a higher scatter in the data. We observed variations in spectral slope between the average spectra from the two nights, which could be due to differences in the phase angle (19.4°) (Sanchez et al. 2012). For our analysis we used the spectrum from Dec. 12$^{th}$, 2008.

The visible spectrum shows a steep rise in reflectance between 0.30-0.41 μm with a much shallower rise between 0.44-0.55 μm and a convex profile beyond that. A shallow absorption band (~3 %) at 0.48 μm cannot be ruled out. Apart from this possible weak feature no distinct absorption band is seen in the entire visible wavelength range. The lower reflectance towards longer wavelength end (~0.98 μm) suggests a possible weak absorption band in the near-IR. However, due to the point-to-point scatter is high in this wavelength region (decreased response of the detector), we are unable to confirm this feature.

*Near-IR Spectrum*

140    The combined visible-NIR spectrum of 2008 EV5 is shown in Fig. 2. The high
141 scatter in the NIR data is due to suboptimal observing conditions. Scatter at 1.1, 1.4, 1.9,
142 and 2.5 µm is due to incomplete correction of telluric water absorption bands. The
143 spectrum is essentially featureless with no evident absorption bands. The lack of
144 absorption features suggests either the presence of a spectrally neutral material (no
145 transition metal bearing minerals) or the masking of absorption bands by opaque material
146 such as carbon or metal (Gaffey et al. 2002). The spectrum (0.75-2.5 µm) has an overall
147 weak negative slope (-0.0785), with increasing scatter towards longer wavelength.
148 *Albedo and Thermal Modeling*
149    Busch et al. (2011) using absolute magnitude $H$ as defined by Pravec and Harris
150 (2007) calculated the optical albedo of 2008 EV5 to be 0.12±0.04. We expect a thermal
151 contribution to the spectrum of 2008 EV5 due to its lower albedo (0.12±0.04) and close
152 proximity to the Sun (0.994 AU) at the time of observation. Using the thermal model
153 described in Reddy, (2009) and Reddy et al. (2012), we modeled (Fig. 3) the upper limit
154 albedo of 2008 EV5 to be 0.20 assuming no thermal emission long ward of 2.2 µm. This
155 thermal model is based on a modified version of the Standard Thermal Model and uses an
156 emissivity value of 0.90 and a beaming parameter 0.756. Due to the scatter in the data we
157 are not formally able to exclude a range of albedos (Fig. 2) with a lower limit at 0.05 for
158 2008 EV5. Hence, the albedo of 0.12±0.04 estimated by Busch et al. (2011) is consistent
159 with our thermal modeling results.
160    Taxonomic classification by Somers et al. (2008) shows that 2008 EV5 is a C-or
161 X type NEA. The mean albedo for C-complex NEAs from Thomas et al. (2011) is
162 0.13±0.06. Therefore, the optical albedo range based on $H$ magnitude estimated by Busch et al. (2011) (0.12±0.04) is consistent with C-type taxonomic class. The primary source of uncertainty here is the $H$ magnitude that is derived from an assumed slope parameter $G$ of 0.15. Slope parameter is dependent on taxonomic type with low albedo asteroids having a $G$ value <0.09 and brighter asteroids having a $G$ value >0.21 (Harris 1989). The appropriate $G$ value for C-type would be 0.086, which would lower the H value and increase the albedo (Harris 1989).

*Meteorite Analogs*

The lack of diagnostic mineral absorption features makes it challenging to identify meteorite analogs for 2008 EV5. Nevertheless, using moderate albedo values 0.12±0.04, negative spectral slope (-0.0785), radar albedo, and circular polarization ratio (SC/OC) from Busch et al. (2011) we can narrow down possible meteorite analogs. Very few meteorites/minerals have featureless spectra with negative spectral slopes. Among the meteorites studied by Gaffey (1976), five meteorites/material have featureless spectra with negative slopes: nickel metal, iron meteorite Juncal, enstatite achondrite Cumberland Falls, CV3 carbonaceous chondrite Grosnaja, and CI carbonaceous chondrite Orgueil.

Busch et al. (2011) note that 2008 EV5 is not metal-rich based on its radar albedo of 0.29±0.09. This excludes metallic nickel or iron meteorites like Juncal as possible analogs for 2008 EV5. Shepard et al. (2010) used radar albedo to constrain the bulk density, metal abundance, and meteorite analogs of main belt asteroids. The 2008 EV5 radar albedo of 0.29±0.09 is within the radar albedo range for CB chondrites (0.20-0.42), stony irons/mesosiderites (0.10-0.28), and EH/EL chondrites (0.10-0.26). However, mesosiderites have red slopes and pyroxene absorption bands unlike 2008 EV5. EH/EL chondrites studied by Gaffey (1976) also have red slopes and some have a weak absorption band at 0.90 μm. So it is unlikely that mesosiderites and EH/EL chondrites are possible analogs for 2008 EV5. CB chondrites have radar albedos (0.20-0.42) that span the range observed for 2008 EV5 (0.29±0.09). CB chondrites are named after Bencubbinites, a class of highly reduced, primitive, brecciated meteorites consisting of (30–60) vol% silicate clasts and (40–70) vol% metal (Weisberg et al., 1990, 2001; Rubin et al., 2003). These silicate clasts are primarily Fe-poor olivine and pyroxene. However, CB chondrites do not have negative slopes as seen in 2008 EV5.

Radar circular polarization ratio is a key indicator of surface roughness and has been correlated to taxonomic types (Benner et al. 2008). E-type asteroids, probable parent bodies of enstatite achondrites (Gaffey et al. 1992), have very high SC/OC (mean 0.892±0.079) (Benner et al. 2008) compared to other probable taxonomic classes for 2008 EV5. 2008 EV5's SC/OC ratio is 0.40±0.07, which makes enstatite achondrites like Cumberland Falls unlikely meteorite analogs. Optical albedos of enstatite achondrites are also very high (mean 0.43) (Mainzer et al. 2011) compared to 2008 EV5 (0.12±0.04). CV3 Carbonaceous chondrite Grosnaja and 2008 EV5 have similar spectral slope. However, the albedo of Grosnoja (0.05) is at the lower limit (Gaffey 1976) of presented 2008 EV5 albedo range, making it less likely to be a possible analog for 2008 EV5.

*CI Chondrite Option:* CI carbonaceous chondrites, such as Orgueil, have albedos (0.05) that are at the lower limit of 2008 EV5's albedo range. Orgueil is spectrally dominated by serpentine and magnetite (Cloutis et al. 2011). The laboratory spectrum of magnetite has a low overall reflectance and a weak absorption band at 0.48 μm due to a spin-forbidden $Fe^{3+}$ absorption band (Figure 4). Cloutis et al. (2011) attribute the blue slope of magnetite spectrum to small grain size. A blue spectral slope is exhibited by several other CI chondrites and also attributed to the small grain size of magnetite (Cloutis et al. 2011). Spectral comparisons of 2008 EV5 to magnetite (Fig. 4) and Orgueil (Fig. 5) show good matches. Variations in spectral slopes and mismatch in the shorter wavelength data could be attributed to differences in phase angle between the lab spectra (30°) and telescopic data (63.2°) (Cloutis et al. 2011).

A comparison of the continuum removed 0.48-µm spin-forbidden $Fe^{3+}$ absorption band (Fig. 6) shows a good match between the telescopic data of 2008 EV5 (gray) and CI chondrite Orgueil (red). A moderately high radar albedo suggests some metal content on 2008 EV5, but CI chondrites have little to none (Cloutis et al. 2011). CR, CO, CV, and CH chondrites all have metal content ranging from 1-20 vol. %. Although metal abundance, optical albedo, and radar albedo of 2008 EV5 are at the higher limits of this meteorite type, we suggest that CI chondrites, such as Orgueil, are plausible analogs for 2008 EV5 based on the spectral match and the presence of the 0.48-µm spin-forbidden $Fe^{3+}$ absorption.

**Summary**

2008 EV5 has a spectrum consistent with the CI carbonaceous chondrite Orgueil. The spectrum is essentially featureless except for a shallow 0.48-µm spin-forbidden $Fe^{3+}$ absorption band. The overall spectrum has a blue slope, which is typical of many CI chondrites. The albedo of 2008 EV5 remains uncertain with a lower limit at 0.05 and higher end at 0.20 based on thermal modeling. The Busch et al. (2011) estimate of 0.12±0.04 is consistent with thermal modeling. Our results are consistent with a C-type taxonomic classification (Somers et al. 2008).

**Acknowledgement**

This research was supported by NASA NEOO Program Grant NNX07AL29G, and NASA Planetary Geology and Geophysics Grant NNX07AP73G. VR would like to thank Javier Licandro for his helpful comments. We thank the IRTF TAC for awarding time to this project, and to the IRTF TOs and MKSS staff for their support.

**Table 1. Observational circumstances for 2008 EV5**

| Wavelength Range | Date of Obs. (UT) | Phase Angle for Asteroid | V. Magnitude |
|---|---|---|---|
| 0.38-0.92 μm | 12/27/2008 | 33.2° | 13.2 |
|  | 01/18/2009 | 52.6° | 16.2 |
| 0.75-2.50 μm | 12/20/2008 | 63.2° | 14.18 |

**Figure Captions**

Figure 1. Visible wavelength spectrum (0.30-0.92 μm; R~500) of 2008 EV5 obtained on December 27[th], 2008, using the dual-channel long-slit CCD spectrometer on the Palomar Mountain 200-inch telescope when the asteroid was 13.2 visual magnitude.

348  Figure 2. Combined Visible (gray) and NIR (black) spectra of 2008 EV5 showing a
349  nearly featureless spectrum with a negative slope. The NIR spectrum (0.75-2.5 µm;
350  R~150) was obtained using the SpeX instrument on NASA IRTF in low-resolution prism
351  mode.
352
353  Figure 3. Continuum removed NIR spectrum of 2008 EV5 along with modeled Planck
354  curves for various albedos. The thermal modeling was accomplished using a modified
355  version of Standard Thermal Model described in Reddy et al. (2012).
356
357  Figure 4. Combined Visible and NIR spectrum of 2008 EV5 along with the spectrum of
358  magnetite (Cloutis et al. 2011). The differences between the two spectra at shorter
359  wavelength could be due to slope variation that is phase angle dependent.
360
361  Figure 5. Combined Visible and NIR spectrum of 2008 EV5 along with the spectrum of
362  CI carbonaceous chondrite Orgueil (Gaffey 1976). The differences between the two
363  spectra at shorter wavelength could be due to slope variation that is phase angle
364  dependent.
365
366  Figure 6. Continuum removed 0.48-µm feature for 2008 EV5 and CI chondrite Orgueil.
367  The feature is likely due to a spin-forbidden $Fe^{3+}$ absorption band (Cloutis et al. 2011).
368  The errors plotted are standard errors of the mean.
369
370

371	**Figure 1. Potentially Hazardous Asteroid 2008 EV5**

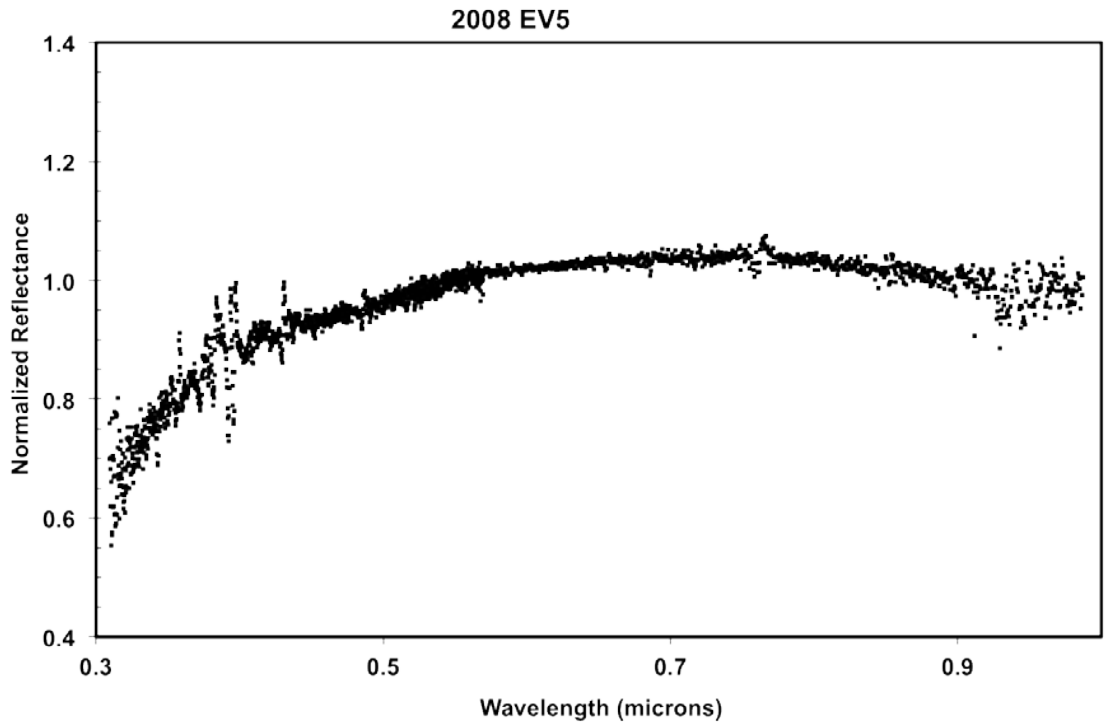

372
373
374
375
376
377
378
379
380
381
382
383

384    **Figure 2. Potentially Hazardous Asteroid 2008 EV5**

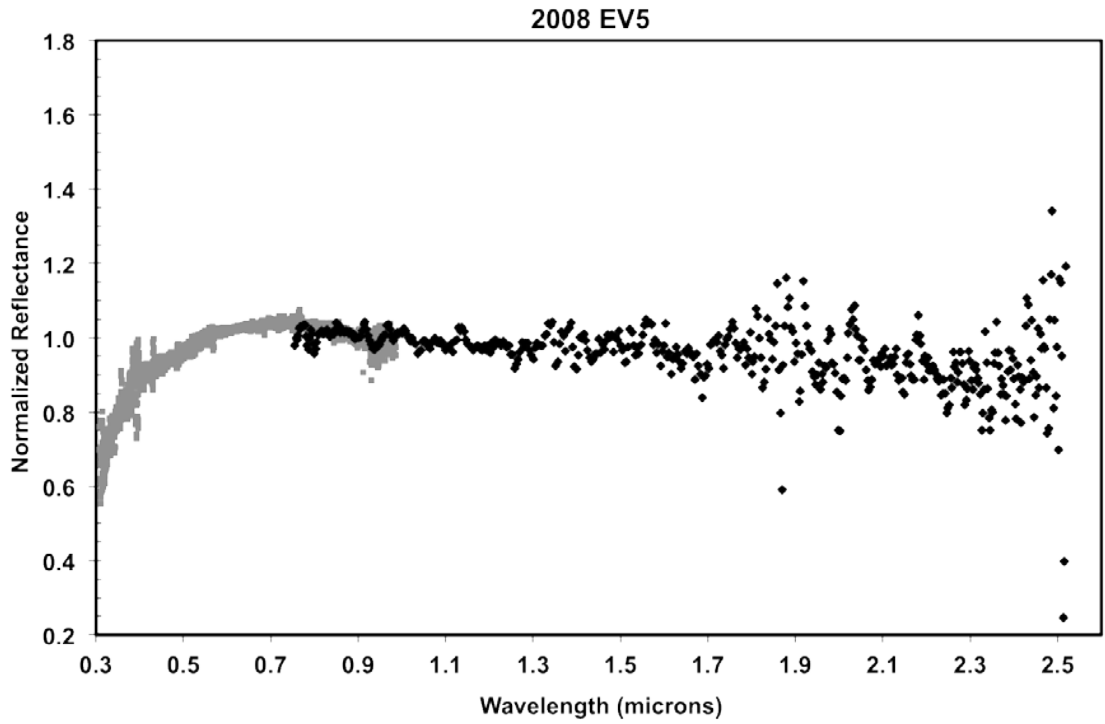

385
386
387
388
389
390
391
392
393
394
395
396

397    **Figure 3. Potentially Hazardous Asteroid 2008 EV5**

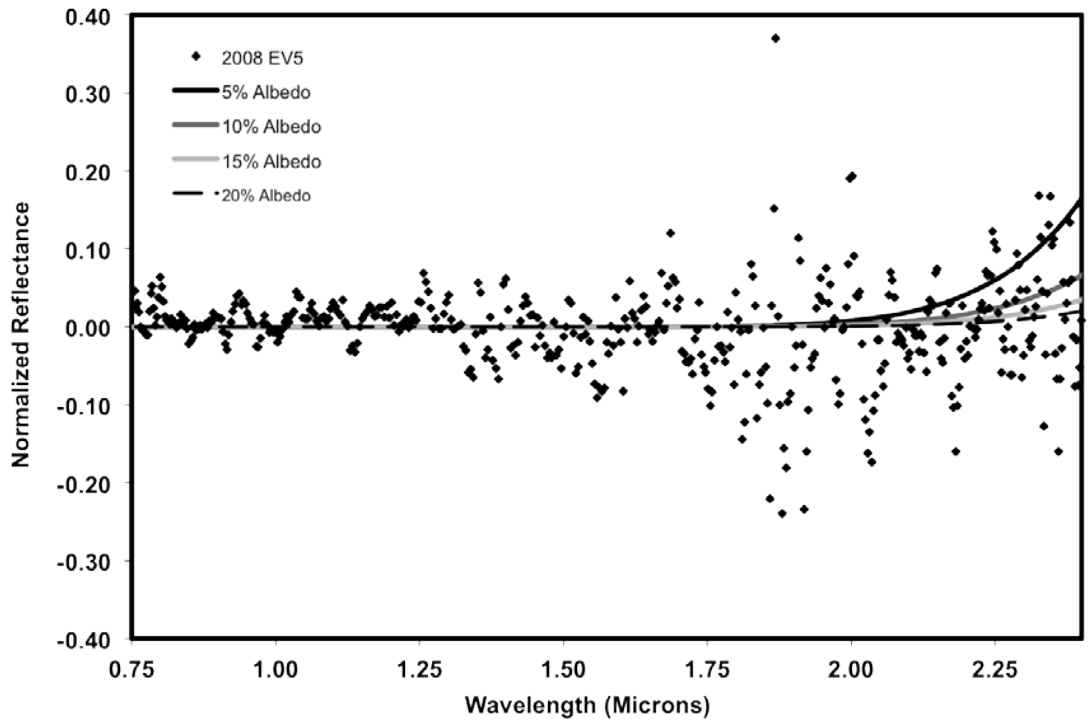

410    **Figure 4. Potentially Hazardous Asteroid 2008 EV5**

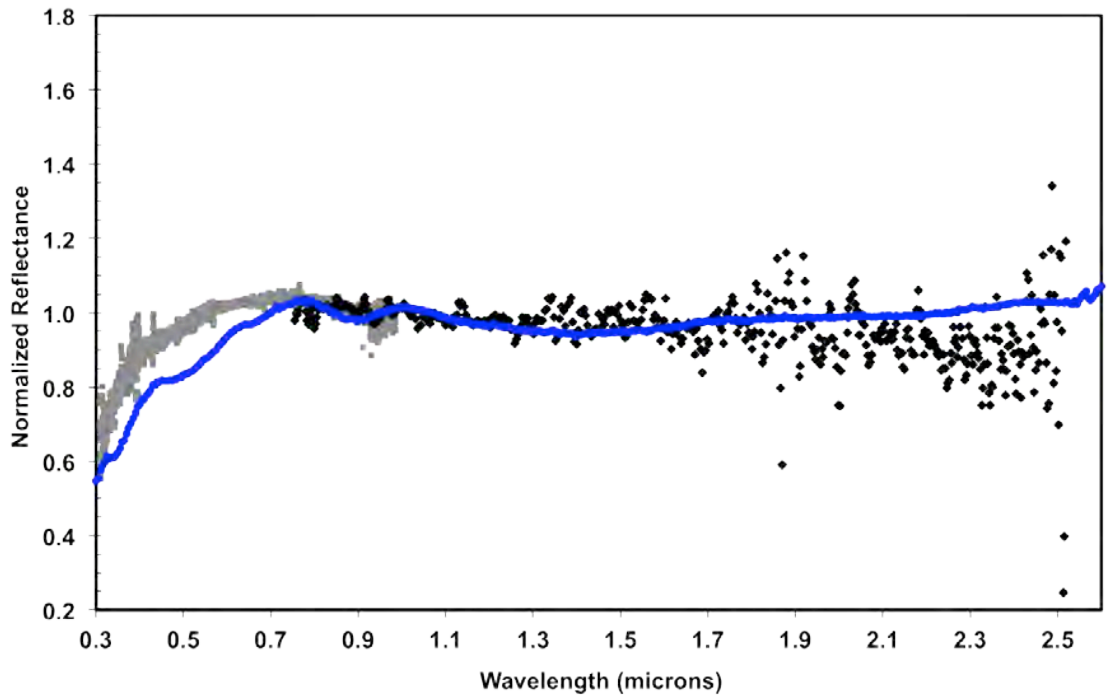

411
412
413
414
415
416
417
418
419
420
421
422

423  **Figure 5. Potentially Hazardous Asteroid 2008 EV5**

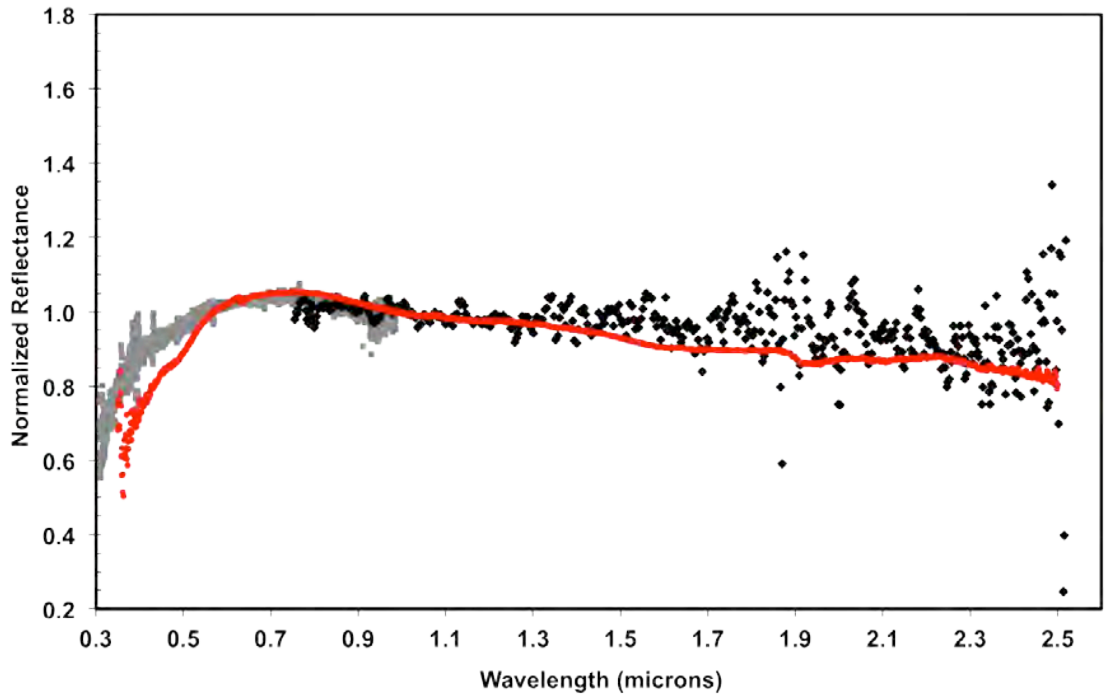

436 **Figure 6. Potentially Hazardous Asteroid 2008 EV5**

437

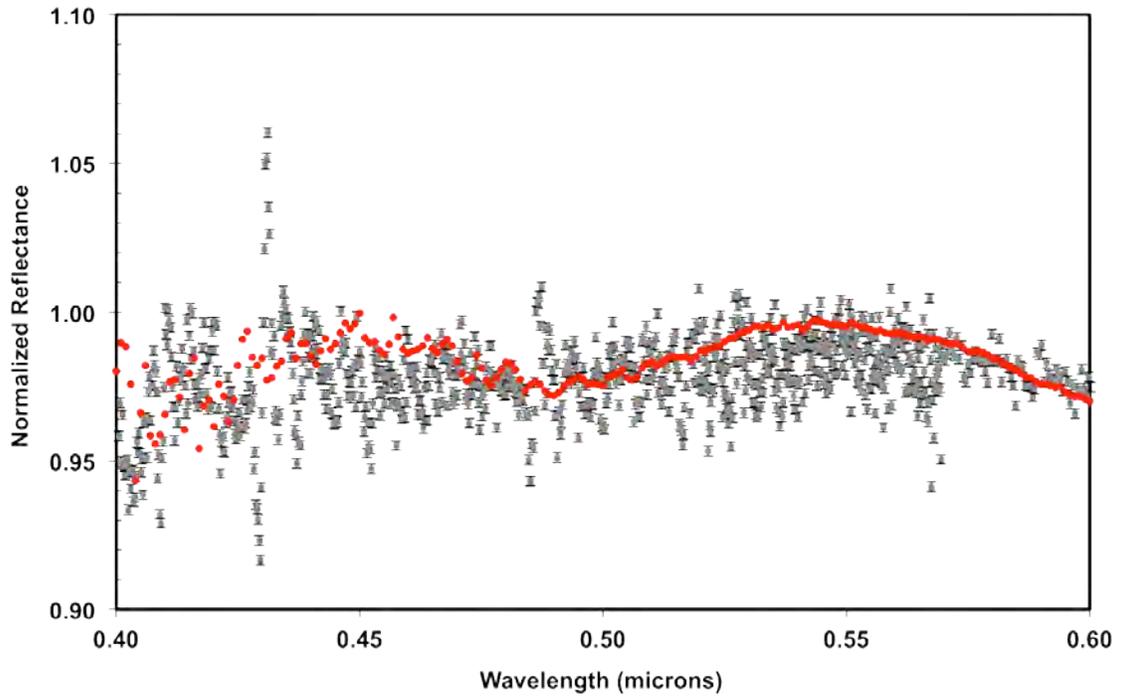

438